\documentclass[12pt]{iopart}
\usepackage{graphicx}
\usepackage{amssymb}

\begin{document}

\title[{\footnotesize Profumo \& Jeltema: Extragalactic IC Light from Dark Matter and the Pamela Positron Excess}]{\mbox{Extragalactic Inverse Compton Light from Dark Matter}\\[0.1cm] Annihilation and the Pamela Positron Excess}

\author{Stefano Profumo}
\address{Department of Physics, University of California, Santa Cruz, CA 95064, USA, and Santa Cruz Institute for Particle Physics, University of California, Santa Cruz, CA 95064, USA}
\ead{profumo@scipp.ucsc.edu}

\author{Tesla E. Jeltema}
\address{UCO/Lick Observatories, Santa Cruz, CA 95064, USA}
\ead{tesla@ucolick.org}

\begin{abstract}
We calculate the extragalactic diffuse emission originating from the up-scattering of cosmic microwave photons by energetic electrons and positrons produced in particle dark matter annihilation events at all redshifts and in all halos. We outline the observational constraints on this emission and we study its dependence on both the particle dark matter model (including the particle mass and its dominant annihilation final state) and on assumptions on structure formation and on the density profile of halos. We find that for low-mass dark matter models, data in the X-ray band provide the most stringent constraints, while the gamma-ray energy range probes models featuring large masses and pair-annihilation rates, and a hard spectrum for the injected electrons and positrons. Specifically, we point out that the all-redshift, all-halo inverse Compton emission from many dark matter models that might provide an explanation to the anomalous positron fraction measured by the Pamela payload severely overproduces the observed extragalactic gamma-ray background.
\end{abstract}

\maketitle

\section{Introduction}

A compelling scenario for the particle nature of the dark matter, which in the standard model of cosmology makes up most of the matter content of the universe, is that of weakly interacting massive particles, or WIMPs \cite{Bergstrom:2009ib}. Predicted in numerous and well-motivated extensions of the minimal (particle physics) Standard Model \cite{Jungman:1995df, Hooper:2007qk, Bertone:2004pz}, WIMPs can naturally account for the cosmological dark matter abundance via standard thermal decoupling from the relativistic thermal bath in the early universe, a mechanism that predicts the relic abundance of electro-weak scale weakly interacting particles to be on the same order of magnitude as the critical density of the universe. Within this framework, in today's cold and clumpy universe WIMPs should still pair annihilate, yielding as stable products energetic particles, including electrons, positrons, nuclei, anti-nuclei and gamma rays, that could provide {\em indirect} evidence for the particle nature of dark matter.

The detection of an increasing positron fraction as a function of energy reported by the Pamela collaboration \cite{Adriani:2008zr} spurred a great deal of interest in the particle physics community as that is in principle one way in which the pair-annihilation of galactic WIMPs can manifest itself. While the ``positron excess'' might be ascribed to astrophysical sources, including as prime candidates nearby mature pulsars \cite{Hooper:2008kg, Yuksel:2008rf, Profumo:2008ms}, the recent high-statistics Fermi-LAT data on the total electron-positron flux \cite{Abdo:2009zk} indicate that galactic WIMP dark matter annihilation is still an open possibility \cite{Grasso:2009ma}, at least for some annihilation final states. In particular, for standard assumptions on galactic cosmic ray production and propagation, a class of dark matter models that give satisfactory fits to the available data appears to be one where the annihilation proceeds into such final states as $\mu^+\mu^-$ or $\tau^+\tau^-$ pairs, or multiple such pairs, with particle dark matter masses of the order of 1-3 TeV \cite{Bergstrom:2009fa, Meade:2009iu}. The required pair annihilation rate for such scenarios are rather large, exceeding by roughly three orders of magnitude what is expected from the above-mentioned thermal decoupling mechanism, i.e. a thermally averaged pair annihilation cross section times relative velocity, at zero temperature, of $\langle\sigma v\rangle=3\times 10^{-26}$ cm$^3$/s. 

While particle physics scenarios have been envisioned that could account for both thermally produced dark matter {\em and} such large annihilation rates today \cite{ArkaniHamed:2008qn, Nomura:2008ru}, it was also pointed out that the annihilation of dark matter at high redshift is rather severely constrained by the resulting injected electromagnetic energy, be it in the form of extragalactic light from the first collapsed structures, or of distortions to the spectrum of the cosmic microwave background \cite{Kamionkowski:2008gj, CyrRacine:2009yn}. In any case, if WIMPs are to explain the galactic positron excess, dark matter annihilation must, by definition, produce energetic (multi-GeV) electrons and positrons. In turn, these light leptons will dominantly loose energy radiating synchrotron light in the presence of magnetic fields and by up-scattering intervening background photons, such as those in the cosmic microwave background (Inverse Compton scattering). Secondary radiation from electrons and positrons produced in dark matter annihilation yields a wide emission spectrum, that spans several decades of the electromagnetic spectrum, from radio frequencies all the way up to gamma rays with energies as large as the mass of the annihilating particle. Indirect dark matter detection through its {\em multi-wavelength} manifestations has been discussed in a number of recent studies, e.g. in Ref.~\cite{gondolo, bertone0101134, aloisio0402588, Bergstrom:2006ny, ullioregis} for the Galaxy, in Ref.~\cite{baltzwai,coladraco,xrdwarf} for local dwarf galaxies and in Ref.~\cite{totani, Colafrancesco:2005ji,ophiuchus,colabullet} for galaxy clusters. The importance of the Inverse Compton secondary emission was also recently emphasized for models that could explain with dark matter annihilation the Pamela positron excess in Ref.~\cite{Pinzke:2009cp} in the context of galaxy clusters, in Ref.~\cite{Meade:2009iu} in the case of the Galaxy, and, very recently, in Ref.~\cite{Ishiwata:2009dk} for decaying dark matter in the Galaxy and at cosmological distances.

With the successful launch and initial science operations of the Fermi Gamma-Ray Space Telescope \cite{glastref, glastnew}, one of the most compelling ways to search for visible effects of WIMP annihilation is via the detection of GeV gamma rays resulting from particle annihilation in high-density regions \cite{Baltz:2008wd}. In particular, given the intrinsic survey-mode functionality of the Fermi Large Area Telescope (LAT), a promising dark matter detection channel relies on the study of the extragalactic gamma-ray flux, with the aim of detecting photons from dark matter annihilation at {\em all red-shifts} and in {\em all dark-matter halos} (i.e. in dark matter halos of any size and at any distance). Pivotal studies and proofs of principle for this technique include Ref.~\cite{BEU, Ullio:2002pj}, that focused on gamma rays produced promptly in the dark matter annihilation event, be it from final state radiation, or from the decay of unstable species resulting in the Standard Model pair-annihilation channel, yielding e.g. $\pi^0\to\gamma\gamma$ from the hadronization chain of a final state quark.

In the present study, we calculate the all-redshift, all-halo emission from the {\em secondary} radiation produced by energetic electrons and positrons ($e^\pm$) via the inverse Compton (IC) scattering of photons in the cosmic microwave background (CMB). The peak of this emission depends on both cosmology (namely, the dark matter halo properties as a function of redshift and halo size) as well as on the $e^\pm$ injection spectrum, and therefore on the particle dark matter mass and its annihilation modes into ``ordinary'' (Standard Model) particles. Generically, the expectation is for the IC emission peak to appear anywhere between soft X-ray frequencies (for light dark matter particles and annihilation final states producing soft $e^\pm$ spectra) and multi-GeV gamma-ray energies (for heavy dark matter candidates pair-annihilating into final states producing a hard $e^\pm$ injection spectrum). Interestingly, the location of the IC peak is roughly redshift independent, since the larger energy of high-redshift target CMB photons is exactly compensated by the redshifting of the IC-emitted photon from production to the observer. Additionally, the IC all-redshift, all-halo emission has the obvious and yet intriguing property that the $\sim (1+z)^4$ suppression in the electron equilibrium number density (driven by a $\sim (1+z)^4$ enhancement of the energy loss rate, dominated by IC and proportional to the CMB energy density) is exactly compensated by the $\sim (1+z)^4$ boost to the resulting IC emission (which is also directly proportional to the energy density of the up-scattered photon field).

We outline in sec.~\ref{sec:calc} the calculation of the all-redshift, all-halo IC emission from $e^\pm$ produced in cosmological WIMP dark matter annihilation, and we present observational constraints in sec.~\ref{sec:data} and our results in sec.~\ref{sec:results}. In that section, we study the emission in detail for several choices of particle dark matter models, and we show that constraints from the X-ray and soft gamma-ray extragalactic background on dark matter annihilation are complementary to those from the gamma-ray energy range probed by Fermi-LAT. We then present perhaps the main upshot of the present study in sec.~\ref{sec:pamela}, where we argue that a wide class of dark matter models that explain the Pamela positron excess and are compatible with the Fermi-LAT $e^\pm$ data overproduce GeV gamma rays via IC emission at all redshifts. 

\section{The all-redshift, all-halo Inverse Compton Light from Dark Matter}\label{sec:calc}

The emission from cosmological dark matter annihilation into gamma-rays promptly produced in the annihilation event has been outlined and calculated in detail in Ref.~\cite{Ullio:2002pj}. We refer the reader to that study for details. The emission can be cast as
\begin{equation}
\frac{{\rm d}\phi_\gamma}{{\rm d}E_0}=\frac{\langle\sigma v\rangle}{8\pi}\frac{c}{H_0}\frac{\bar\rho^2_m}{m^2}\int\ {\rm d}z\frac{(1+z)^3\Delta^2(z)}{h(z)}\frac{{\rm d}N_{\rm GR}}{{\rm d}E}\left(E=E_0(1+z)\right),
\end{equation}
where ${\rm d}N_{\rm GR}/{\rm d}E$ indicates the differential number of gamma rays promptly produced in the annihilation of the mass $m$ WIMP, $\Delta^2(z)$ is the redshift-dependent factor that encompasses the effect of structure in the dark matter distribution (we will comment below on how this factor depends on assumptions on the redshift dependence of halo concentration and on the assumed halo density profile), $\bar\rho_m$ is today's average dark matter density, $c$ is the speed of light, $H_0$ is the Hubble parameter, and 
\begin{equation}
h(z)=\sqrt{\Omega_m(1+z)^3+\Omega_\Lambda},
\end{equation}
where we fix here the cosmological parameters to $\Omega_\Lambda=0.701$ and $\Omega_m=1-\Omega_\Lambda$. In the case of the IC emission,  ${\rm d}N_{\rm GR}/{\rm d}E$ is replaced by the (redshift-dependent) IC emission function per WIMP annihilation
\begin{equation}
\frac{{\rm d}N_{\rm IC}}{{\rm d}E}(E,z)=\int {\rm d}E_e\ \frac{{\rm d}\tilde n_e}{{\rm d}E}(E_e,z)\ W_{\rm IC}\left(E,E_e,z\right),
\end{equation}
where
\begin{equation}
W_{\rm IC}\left(E,E_e,z\right)=c\int{\rmd}\varepsilon\ n_\gamma(\varepsilon,z)\sigma_{\rm KN}(E,E_e,\varepsilon),
\end{equation}
(in the equation above $n_\gamma(\varepsilon,z)$ indicates the cosmic microwave background photon spectrum at redshift $z$ and energy $\varepsilon$, and $\sigma_{\rm KN}$ is the differential Klein-Nishina cross section formula; for further details see Ref.~\cite{Colafrancesco:2005ji}), and where we indicate with
\begin{equation}\label{eq:eqdist}
\frac{{\rm d}\tilde n_e}{{\rm d}E}(E_e,z)=\frac{1}{b(E_e,z)}\int_{E_e}^m {\rm d}E^\prime\ \frac{{\rm d}N_{\rm e}}{{\rm d}E}(E^\prime).
\end{equation}
In the Equation above, ${\rm d}N_{\rm e}/{\rm d}E$ stands for the differential number of electrons plus positrons produced in a WIMP annihilation event, and 
\begin{equation}
b(E,z)\approx2.67\times10^{-17}(1+z)^4(E/{\rm GeV})^2\ {\rm GeV}/{\rm s}
\end{equation}
is the energy loss rate, dominated, at high redshifts, by the IC scattering off of CMB photons. Notice that in Eq.~(\ref{eq:eqdist}) we (i) neglect diffusion, which is clearly irrelevant for the computation of the cosmological flux under consideration here, and (ii) we assume that electrons and positrons loose energy instantaneously, and therefore reach an equilibrium distribution at the same redshift at which they emit the IC radiation. This second approximation depends on the fact that at all $e^\pm$ energies and redshifts we consider here, the IC energy loss time scale, roughly $E_e/b(E_e)$, is much shorter than the Hubble time at corresponding redshifts. It is therefore legitimate to neglect the effect of the expansion of the universe from the annihilation event that produces the high-energy $e^\pm$ to when these particles have reached the equiblibrium configuration of Eq.~(\ref{eq:eqdist}) and have IC-upscatteredd the intervening CMB photons.

We neglect here synchrotron energy losses, whose contribution can be estimated as
\begin{equation}
b_{\rm sync}(E,z)\approx0.254\times10^{-17}\left(\frac{B}{1\ \mu{\rm G}}\right)^2(E/{\rm GeV})^2\ {\rm GeV}/{\rm s}.
\end{equation}
In principle, the average magnetic field is a function of redshift. Assuming $B$ does not rapidly increase with redshift, synchrotron losses are entirely negligible at large redshift, say $z\gtrsim2$. Taking as a nominal value for $B$ the average magnetic field observed in clusters of galaxies, $B\sim1\ \mu{\rm G}$, even at $z\sim0$ synchrotron losses are sub-dominant with respect to inverse Compton. Although most dark matter annihilation events will occur in high-density regions, the overall average magnetic field might be even smaller than its values in clusters of galaxies, and approach the extremely small values inferred for inter-galactic magnetic fields, $B\lesssim1$ nG. 

In the calculation of the $\Delta^2(z)$ function we consider two extreme possibilities for the dark matter halo density profiles (which we assume in all cases to have a universal profile over the entire halo mass range under consideration), namely the centrally steep Moore profile $g_{\rm Moore}(x)\propto x^{-1.5}(1+x^{1.5})^{-1}$ of Ref.~\cite{moore}, and the centrally cored Burkert profile $g_{\rm Burkert}\propto [(1+x)(1+x^2)]^{-1}$ of Ref.~\cite{burkert}, where $x$ indicates the radial distance from the center of the halo in units of the halo scale length. The Moore profile assumption leads to large central density squared in all halos, and therefore to a large signal compared to the Burkert profile. The widely employed setup of the Navarro-Frenk-White universal halo profile of Ref.~\cite{NFW} falls in between the two cases we consider here.

We also attempt to bracket the uncertainty in the structure formation history by considering two alternative schemes for the redshift and halo-mass dependence of the halo concentration parameter described in Ref.~\cite{Ullio:2002pj}: specifically, we consider the Bullock et al model of Ref.~\cite{bullock} and the Eke, Navarro and Steinmetz (ENS) scenario of Ref.~\cite{ENS}. For both models, we assume a fixed logarithmic interval cutoff for the smallest structure where the model predictions for the concentration parameter are extrapolated, and a cutoff scale of $10^5M_\odot$ at $z=0$ (see the discussion on this point in sec.IV of Ref.~\cite{Ullio:2002pj}). We find that this choice (as compared e.g. with the choice of a redshift-independent small-scale cutoff) is however not critical. The Bullock et al setup leads to (i) a steeper dependence of the concentration parameter on the halo mass (see fig.~2 of Ref.~\cite{Ullio:2002pj}), (ii) weighs more heavily structures at lower redshift and, in general, (iii) always produces a larger prediction than the ENS setup for the total extragalactic light from dark matter annihilation. Our reference setup in sec.~\ref{sec:results} employs the Bullock et al scheme and a universal Moore profile.

An important fact for the present calculation (particularly for dark matter models with large masses) is that the universe is not always transparent to high-energy radiation produced at high redshift. A prominent effect is high energy electron-positron pair-production and gamma rays produced at high redshift impinging on the intervening extragalactic background light (see e.g.~\cite{gi09}). We model this effect with the approximate exponential form of Ref.~\cite{BEU}, which in turn relies on the results presented in Ref.~\cite{somer}. At lower energies other processes also contribute to photon absorption (but where negligible in the energy ranges considered by \cite{BEU} and \cite{Ullio:2002pj}), including photoionization and Compton scattering. We model the photon transparency window according to a numerical interpolation of the results of fig.~2 of Ref.~\cite{chen}.
 
As we remarked in the Introduction, the location in energy $E^{\rm IC}_{\gamma}$ of the peak of the IC emission off of an impinging $e^\pm$ with an initial energy $E_{e^\pm}$, which can be approximated as \cite{longair} 
\begin{equation}
E^{\rm IC}_{\gamma}\approx E_\gamma\left(E_{e^\pm}/m_e\right)^2,
\end{equation}
is approximately redshift-independent. In fact, indicating with $E^{\rm CMB}_{\gamma}$ the average energy of a CMB photon today, the IC up-scattered CMB photons at a redshift $z$  have an energy  $E_\gamma=(1+z)E^{\rm CMB}_{\gamma}$. Those upscattered photons, though, will redshift from the redshift of production to $z=0$ by exactly the inverse factor $1/(1+z)$. Therefore, the IC peak is roughly redshift-independent (the actual shape of the emission at a given redshift is however affected by absorption and by the spectral shape of CMB photons at higher redshifts).

As also pointed out above, there is no suppression of the IC emission at high redshift, since the $\sim(1+z)^4$ increase in the energy losses, dominantly driven by IC scattering of CMB photons, is exactly compensated by an increase in the target photon energy density with redshift, and therefore an increase in the IC emission (which is directly proportional to the energy density of target photons \cite{longair}). Notice that this is unlike the case of the all-redshift and all-halo {\em radio} emission \cite{radio}, where there is no reason to expect that the effective radiation energy density in magnetic fields scales as rapidly with redshift as the CMB energy density. In turn, as remarked in Ref.~\cite{radio}, this implies an effective suppression of any contribution at redshifts $z\gtrsim1$ for the extragalactic {\em radio} emission from dark matter annihilation.

\section{Overview of observational data}\label{sec:data}

We will consider below the high energy extragalactic background at energies ranging from 1 keV, below which the X-ray background is dominated by galactic emission (see e.g. Ref.~\cite{hi06}), through the EGRET measurements of the extragalactic gamma-ray background, which extends to energies as large as roughly 20 GeV.  The soft X-ray background (below about 10 keV) has been studied extensively, including recent measurements with \textit{Chandra}, \textit{XMM}, and \textit{Swift} (see e.g.~\cite{hi06,de04,mo09}).  At soft X-ray energies the extragalactic background has been almost entirely resolved (to the $\sim 80$\% level) into discrete X-ray sources, primarily Active Galactic Nuclei (AGN, see e.g.~\cite{br05,hi06}).  Recent work has additionally shown that the remaining X-ray background at energies below $\sim8$ keV can be significantly reduced, and within the errors entirely accounted for, by removing optical and IR sources detected by \textit{HST} and \textit{Spitzer} \cite{wo06,hi07}.  Here we will use the $2 \sigma$ upper limit on the remaining soft X-ray background from the work of Ref.~\cite{hi07}.

The X-ray background peaks at somewhat higher energies, around 30 keV \cite{gr99,ch07,aj08}, where much of the background remains unresolved by current instruments.  However, modeling based on the observed spectra and X-ray luminosity function at softer energies shows that unabsorbed and Compton-thin AGN account for roughly 75\% of the X-ray background at the peak \cite{gi07,tr09}.  The remainder of the X-ray background can be explained through the contribution of Compton-thick AGN \cite{gi07,tr09}, but the modeling of this component depends heavily on uncertainties in the fraction of Compton-thick AGN and their spectra \cite{tr09}.  In the present work, we will neglect the contribution of Compton-thick AGN, but subtract the model for the contribution of unabsorbed and Compton-thin AGN described in Ref.~\cite{gi07}\footnote{This model is available via the web at http://www.bo.astro.it/$\sim$gilli/xrb.html} to the X-ray background above 10 keV.  We consider this to be a conservative upper limit on the unaccounted-for X-ray background.  In fig.~\ref{fig:finalstate} and following, we show a recent measurement of the hard X-ray background with \textit{Swift} BAT as well as the $2 \sigma$ upper limits once the model of the AGN contribution is subtracted off.  The \textit{Swift} measurements of the X-ray background are consistent with recent \textit{INTEGRAL} measurements \cite{ch07} but about 8\% higher than the previous \textit{HEAO-1} measurements \cite{gr99}.

The contribution of the previously mentioned AGN populations is expected to be negligible above a few hundred keV \cite{gi07,tr09,aj09}.  However, a significant extragalactic background is observed in the MeV \cite{we00, wa00} and GeV range \cite{sr98,st04}.  Here blazars may make a significant contribution and under reasonable assumptions can account for the entire MeV background \cite{aj09}.  However, uncertainties in blazar spectra significantly affect the modeling of this component \cite{ve07,aj09}. Recent studies, see e.g. Ref.~\cite{Inoue:2008pk}, imply that models exist where up to 80\% of the extra-galactic gamma-ray background can be accounted for by blazars and by non-blazar AGNs. In this scenario, most of the blazar sources are predicted to be resolved by Fermi-LAT \cite{Inoue:2008pk}. Here, however, we will conservatively consider the entire background in the MeV and GeV range and simply (and conservatively) require that the dark matter annihilation emission not overproduce the total observed extragalactic gamma-ray background.

\section{Results and Discussion}\label{sec:results}

\begin{figure}[t]
\begin{center}
\mbox{\hspace*{-0.5cm}\includegraphics[width=8.cm,clip]{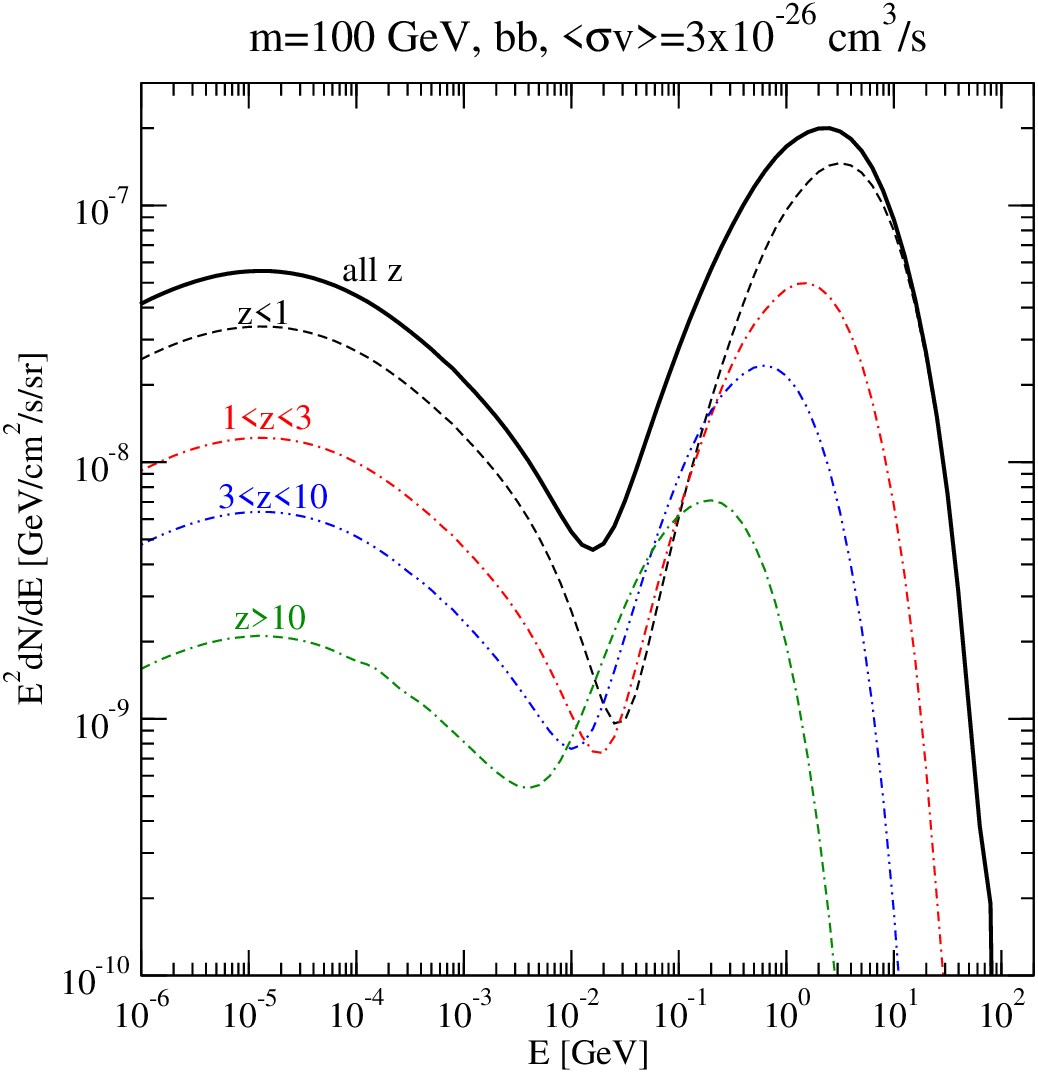} \qquad \includegraphics[width=8.cm,clip]{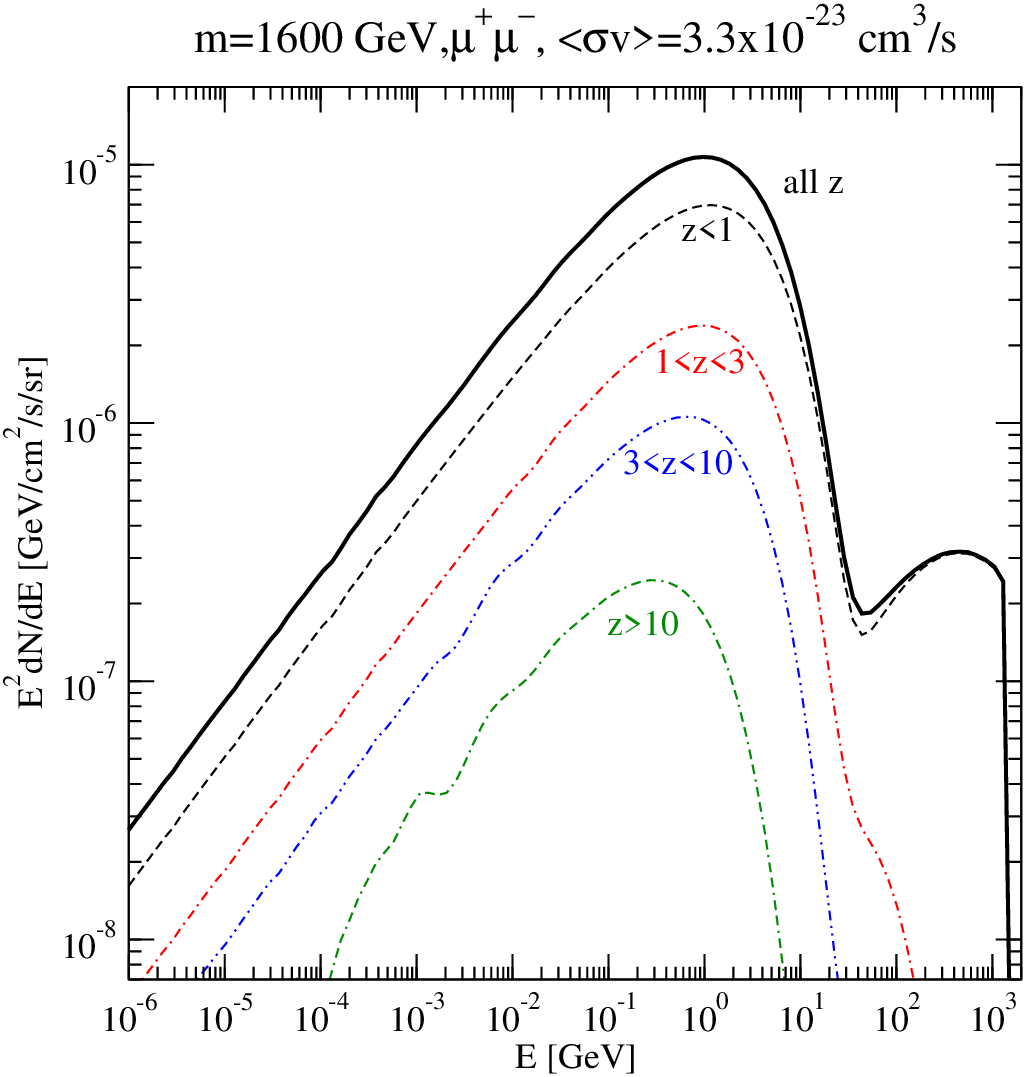}}
\caption{Break-up, in redshift bands, of the total extragalactic emission from dark matter annihilation at all redshifts. The left panel refers to a dark matter particle model pair-annihilating into a $b\bar b$ pair, and weighing 100 GeV, with a standard pair-annihilation rate $\langle\sigma v\rangle=3\times 10^{-26}$ cm$^3$/s. The right panel assumes a 1.6 TeV dark matter particle, with an annihilation rate of  $\langle\sigma v\rangle=3.3\times 10^{-23}$ cm$^3$/s and a dominant $\mu^+\mu^-$ annihilation final state. This setup has been shown in Ref.~\cite{Bergstrom:2009fa} to reproduce (with customary assumptions on the galactic energy losses and the astrophysical positron background) the anomalous positron fraction reported by Pamela \cite{Adriani:2008zr}, as well as the Fermi electron-positron data \cite{Abdo:2009zk}. \label{fig:redshift}}
\end{center}
\end{figure}

\begin{figure}[t]
\begin{center}
\mbox{\hspace*{-0.5cm}\includegraphics[width=8.cm,clip]{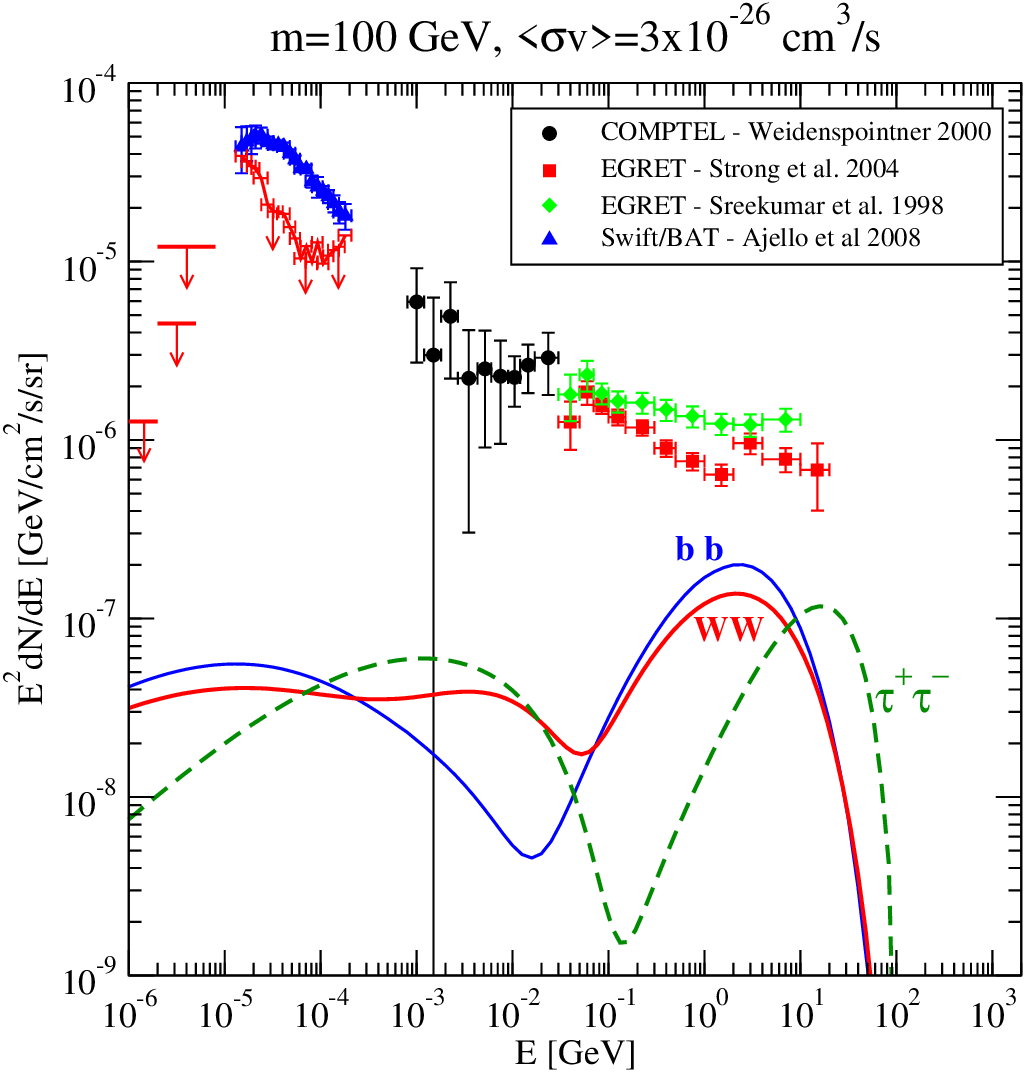} \qquad \includegraphics[width=8.cm,clip]{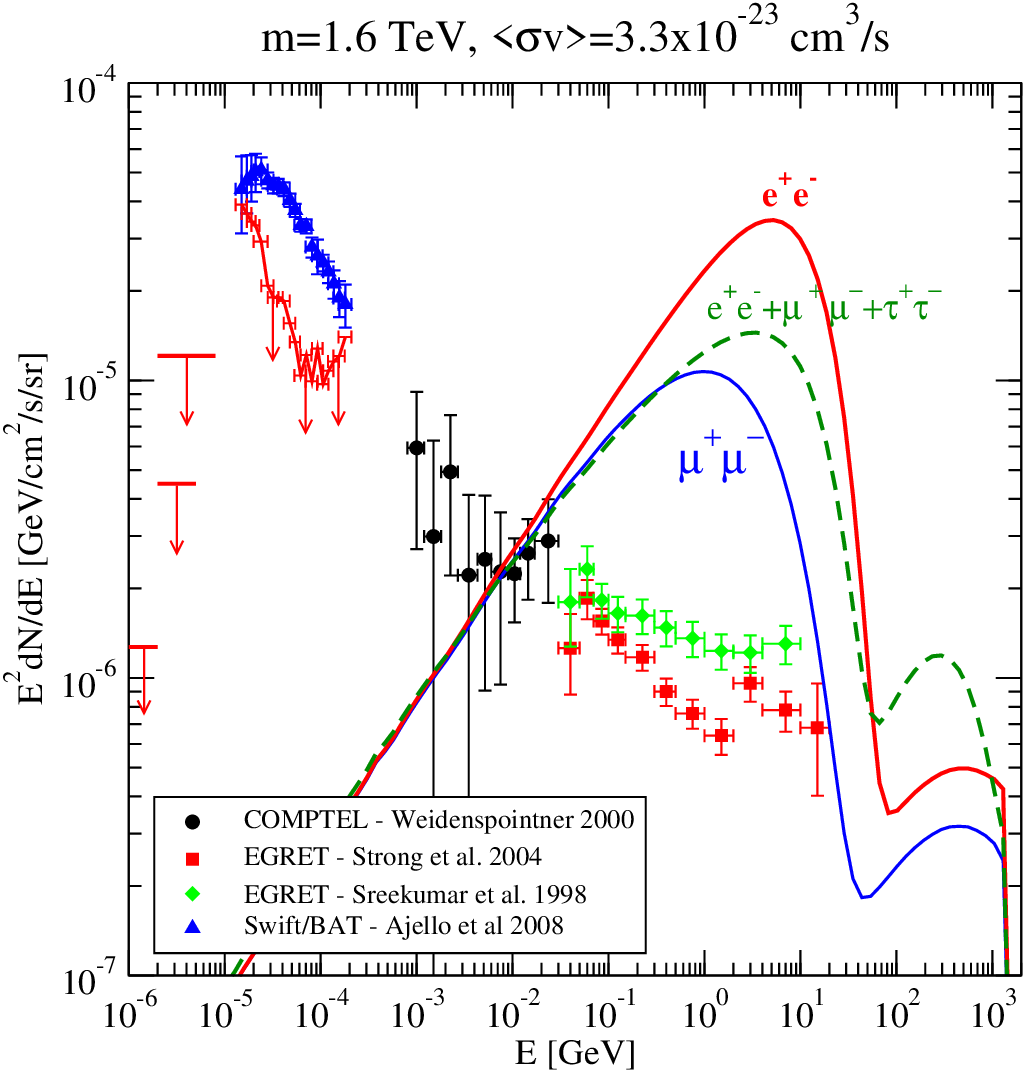}}
\caption{The dependence of the dark matter all-redshift all-halo annihilation emission on the dominant annihilation final state: $b\bar b$, $W^+W^-$ and $\tau^+\tau^-$ in the left panel (all particle models with a pair annihilation cross section $\langle\sigma v\rangle=3\times 10^{-26}$ cm$^3$/s and a mass of 100 GeV). In the right panel we show the final states $e^+e^-$, $\mu^+\mu^-$ and a ``democratic lepto-philic'' setup (equal probability for each charged lepton pair), with a common mass of 1.6 TeV and pair annihilation cross section $\langle\sigma v\rangle=3.3\times 10^{-23}$ cm$^3$/s. The data shown are from Ref.~\cite{hi07} (\textit{Chandra}), \cite{aj08} (\textit{Swift}), \cite{we00} (\textit{Comptel}), \cite{sr98,st04} (EGRET), and the model for hard X-ray AGN is from Ref.~\cite{gi07}.
\label{fig:finalstate}}
\end{center}
\end{figure}

\begin{figure}[t]
\begin{center}
\mbox{\hspace*{-0.5cm}\includegraphics[width=8.cm,clip]{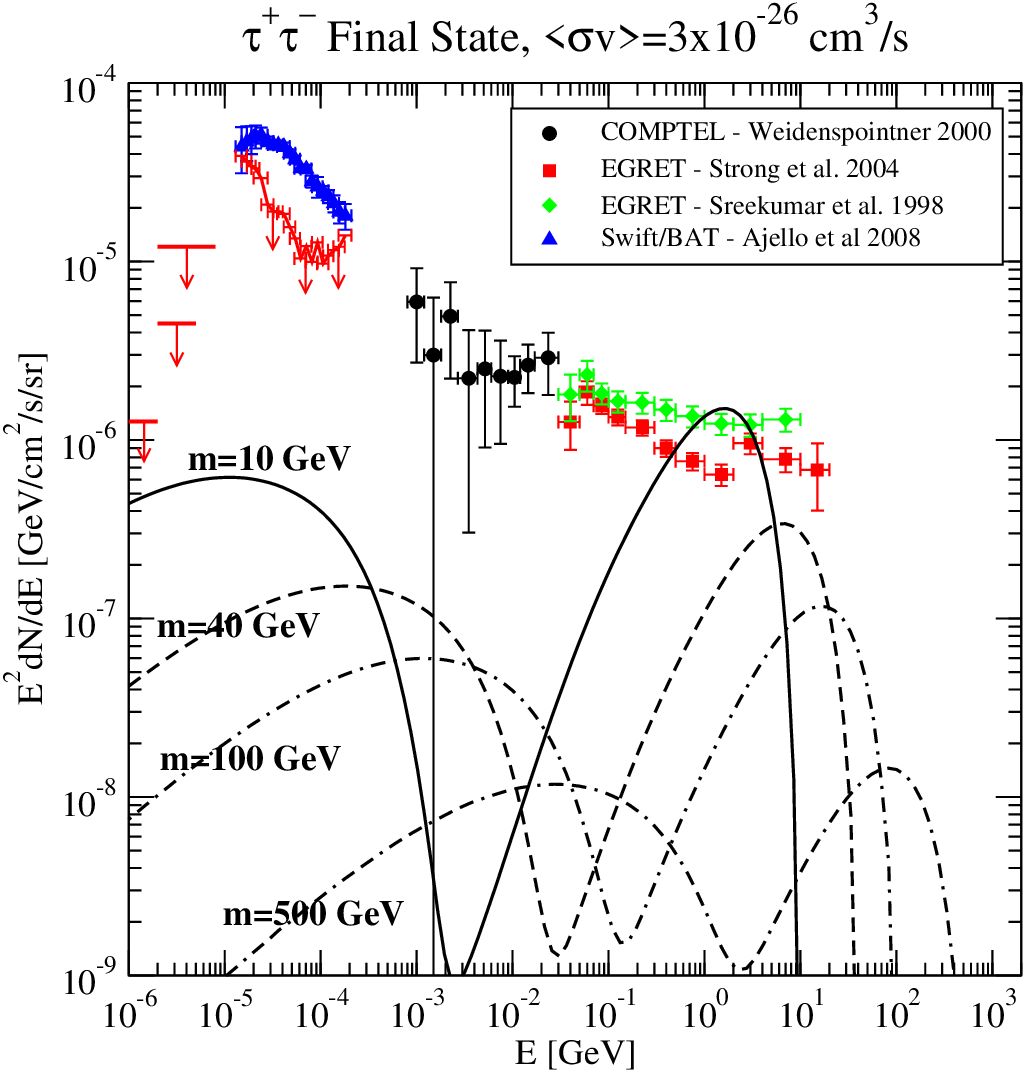} \qquad \includegraphics[width=8.cm,clip]{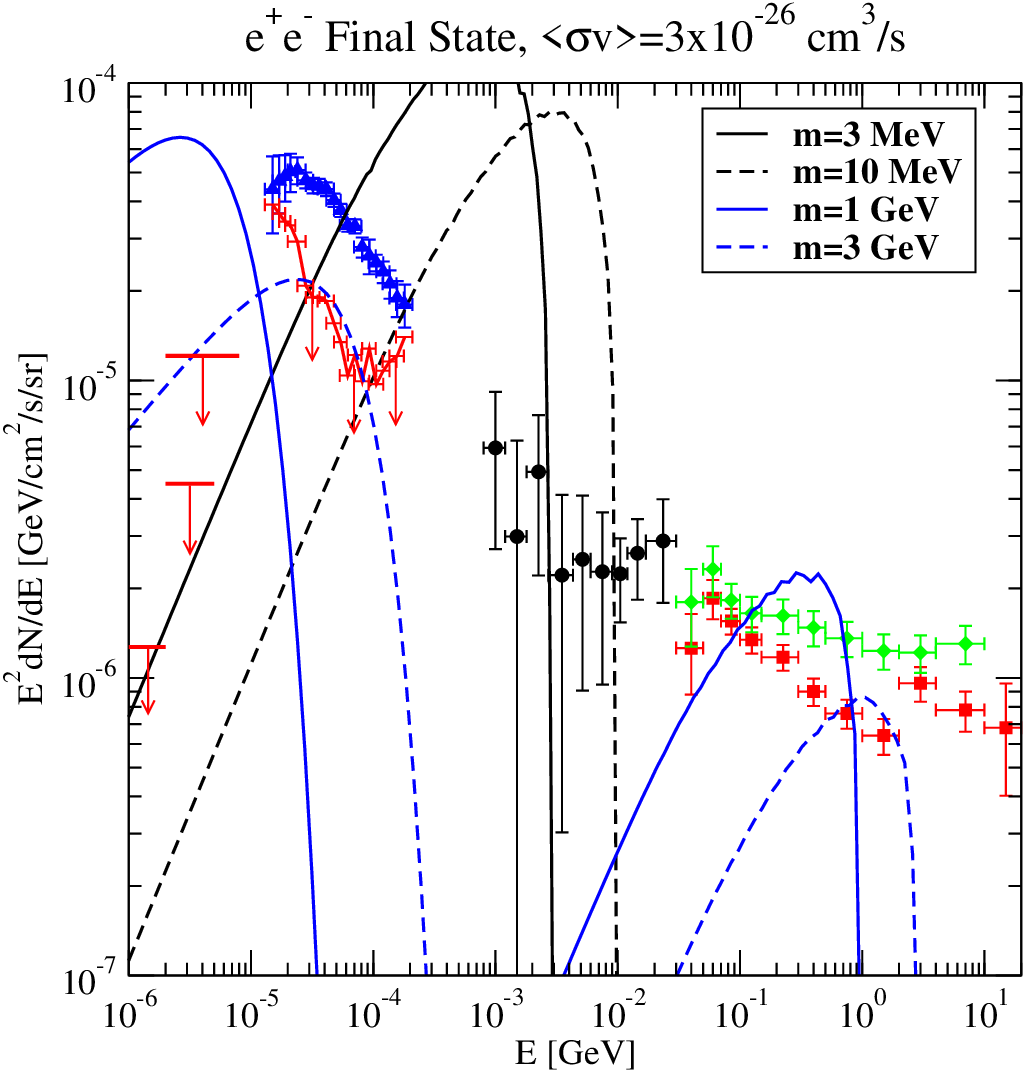}}
\caption{The dependence of the dark matter all-redshift all-halo annihilation emission on the dark matter particle mass. In both panels the pair-annihilation rate is set to $\langle\sigma v\rangle=3\times 10^{-26}$ cm$^3$/s. In the left panel we show a $\tau^+\tau^-$ annihilation final state, for dark matter masses of 10, 40, 100 and 500 GeV, while in the right panel we consider a pure, monochromatic $e^+e^-$ final state featuring dark matter particle masses of 3 MeV, 10 MeV, 1 GeV and 3 GeV. The data shown are as in fig.~\ref{fig:finalstate}\label{fig:mass}}
\end{center}
\end{figure}

In this section we articulate the relevant physical features of the extragalactic all-redshift, all-halo emission from dark matter annihilation, including how different redshift bands contribute to the overall emission (fig.~\ref{fig:redshift}), the dependence on the dominant dark matter pair-annihilation final state (fig.~\ref{fig:finalstate}) and on the particle mass (fig.~\ref{fig:mass}). We also show the corresponding gamma-ray emission promptly produced in the final state.  The presentation employes a bimodal set of dark matter models: in the left panels we show particle models with customary masses, pair annihilation rates and annihilation final states, where ``customary'' alludes to what would be expected in well-motivated theoretical setups beyond the Standard Model, including, but not limited to, supersymmetry \cite{Jungman:1995df} and Universal Extra-Dimensional models \cite{Hooper:2007qk}. In the right panels, instead, we employ models such as those we mentioned in the Introduction that provide a satisfactory fit to the reported Pamela positron excess \cite{Adriani:2008zr} as well as to the null result in the search for an antiproton excess \cite{pamelaantiproton} and to the  $e^\pm$ spectrum measured by Fermi-LAT \cite{Abdo:2009zk}. This latter category features large particle masses, leptonic-dominated annihilation final states, and very large pair-annihilation rates. We remind the reader that in fig.~\ref{fig:redshift}-\ref{fig:mass} we employ the Bullock et al structure formation setup, as specified in the sec.~\ref{sec:calc}, and a universal Moore profile for all dark matter halos.

We show in fig.~\ref{fig:redshift} the contribution of various redshift intervals to the overall all-redshift, all-halo extragalactic emission from dark matter annihilation, including the IC emission described in sec.~\ref{sec:calc}. The dashed black line indicates contributions from $z<1$, the dot-dashed red line from the interval $1<z<3$, the double-dotted-dashed blue line from $2<z<10$ and, finally, the dot-double-dashed green line from $z>10$. For the left panel we employ a model with a mass of 100 GeV annihilating into a $b\bar b$ pair with a standard pair-annihilation rate as preferred by thermal production of dark matter of $\langle\sigma v\rangle=3\times 10^{-26}$ cm$^3$/s. The right panel features a model similar to one of those considered in Ref.~\cite{Bergstrom:2009fa}, with a 1.6 TeV particle annihilating into $\mu^+\mu^-$ pairs at a rate of $\langle\sigma v\rangle=3.3\times 10^{-23}$ cm$^3$/s, equivalent to an effective ``boost factor'' of 1,100.

First, we notice that the break-up in redshift bands allows one to appreciate that while the gamma rays promptly produced in the final state (we shall indicate these as FSGR, final state gamma rays) produce a bump that, as expected, red-shifts to lower and lower energies the higher the redshift at which the annihilation occurs, this, for the reasons explained above in sec.~\ref{sec:calc} is not the case for the IC emission bump, that always sits at $E_\gamma\sim10$ keV in the left panel and at $E_\gamma\sim1$ GeV in the right panel (the location of the IC bump reflecting the very different $e^\pm$ injection spectrum in the two cases). Secondly, we notice that annihilations at intermediate redshifts $1<z<10$ contribute to the overall IC emission almost as much as those at low redshift $z<1$. Even at high redshift, $z>10$, where the contribution to the FSGR emission is negligible at the peak energy, in the IC case we find a contribution which just below $\sim10\%$ of the total. Finally, in the high-mass case shown in the right panel it is clear that the contribution of annihilations at $z>1$ is entirely negligible to the FSGR, due to gamma-ray absorption on the extra-galactic background light, while annihilations at those redshifts significantly contribute to the IC emission.

In fig.~\ref{fig:finalstate} we compare the predictions for the all-redshifts, all-halo dark matter annihilation emission, for different pair annihilation final states, with data on the extragalactic diffuse light (see sec.~\ref{sec:data}). In the left panel we fix the mass at 100 GeV, the pair annihilation at $\langle\sigma v\rangle=3\times 10^{-26}$ cm$^3$/s, and consider the $b\bar b$ (solid blue line), $W^+W^-$ (solid red line) and $\tau^+\tau^-$ (dashed green line) final states. The different injection spectra for these three final states drive the differences in the IC emission: 
\begin{itemize}
\item in the $b\bar b$ case, the $e^\pm$ are almost solely produced by the decays of charged pions produced in the jet resulting from the high-energy final state heavy quark; also, most of the FSGR are similarly produced by neutral pion decays
\item in the $W^+W^-$ case, hadronic $W$ decay modes also yield a population of soft charged pions that produces the same bump (although with a relatively suppressed intensity) at $E_\gamma\sim10$ keV as in the $b\bar b$ case. In addition to this bump, though, a second bump emerges at much higher energies, $E_\gamma\sim10$ MeV, that originates from the {\em leptonic} decay modes of the $W$, yielding energetic $e^\pm$ (for instance promptly from $W\to e\nu_e$ or from the subsequent decays into $e^\pm$ of muons and taus).
\item the $\tau^+\tau^-$ final state produces an IC feature that results from $e^\pm$ injected from both the hadronic and (dominantly) from the leptonic decay of the $\tau$ into $\mu$ subsequently decaying into $e$. 
\end{itemize}
The left panel illustrates that for {\em vanilla} dark matter, the extragalactic background light is best constrained through FSGR rather than in the IC band, although an improvement of the understanding of the extragalactic background in the MeV range could potentially yield complementary information. With the optimistic structure formation and halo model setup we employ here, if Fermi-LAT resolves enough blazars to bring the diffuse extragalactic gamma-ray background down by an order of magnitude, this will start to put very significant constraints on this dark matter detection channel. The soft X-ray band is also potentially interesting, especially for annihilation final states yielding a soft $e^\pm$ injection spectrum. The needed improvement, though, exceeds one order of magnitude for customary dark matter models.

The right panel assume dark matter models with large masses and pair-annihilation rates (respectively, $m=1600$ GeV and $\langle\sigma v\rangle=3.3\times 10^{-23}$ cm$^3$/s), that annihilate into leptonic final states. This category of models is motivated by a new physics interpretation of the positron excess in the Pamela data, and by compatibility with results from various other cosmic ray experiments, including the Fermi-LAT measurement of the $e^\pm$ spectrum. While the red line indicates an $e^+e^-$ final state, the blue line indicates the $\mu^+\mu^-$ case and the dashed green line a democratic lepto-philic case \cite{leptophilic}, where we assume an equal probability of annihilation into any one charged lepton pair. The FSGR in the electron and muon case only arise from the internal bremsstrahlung off of the final state charged leptons, while a contribution from neutral pion decay is present in the $\tau$ case. The location of the IC peak, in the case of the monochromatic $e^+e^-$ annihilation mode, has an obvious location at $E_\gamma\sim E^{\rm CMB}_\gamma(m_{\rm DM}/m_e)^2\sim$few GeV. The peak is at slightly lower energies (and is a bit wider) for the case of muons, featuring a non-monochromatic and lower energy final state electron-positron population. The democratic case interpolates between the above cases. 

The common, striking feature to all the models we show in the right panel is that, in the context of the structure formation and halo setup we employ here, the {\em models compatible with Pamela are ruled out by the IC extragalactic all-redshift and all-halo emission from dark matter annihilation}. The EGRET data \cite{sr98, st04} are well below, by more than one order of magnitude, the predicted extragalactic emission. Fermi-LAT, resolving a much larger number of extragalactic point sources, will likely significantly improve our understanding of the origin of the diffuse extragalactic light at gamma-ray frequencies, and probably tighten further the constraints on dark matter models such as those shown in the right panel as possible candidates to explain the positron anomaly. We get back to this point in sec.~\ref{sec:pamela} and discuss how this conclusion is affected by changing our assumptions on the structure formation and halo model setups.

The energy scale of the  $e^\pm$ injected by dark matter annihilation is set by the dark matter mass. In fig.~\ref{fig:mass} we study the effect of varying the dark matter particle mass in the all-redshift all-halo emission spectrum from IC. In the left panel, we set $\langle\sigma v\rangle=3\times 10^{-26}$ cm$^3$/s, and we assume a dominant annihilation mode into $\tau^+\tau^-$. The black lines show masses corresponding to 10 GeV (solid), 40 GeV (dashed), 100 GeV (dot-dashed) and 500 GeV (dot-double-dashed). Dark matter particles as light as a fraction of a GeV are viable for instance in the context of modified cosmologies, and compatible with supersymmetry (see e.g.~\cite{lightneut}). The left panel illustrates the complementarity of measurements (and of modeling) of the soft (and hard) X-ray diffuse extragalactic background with those at higher frequency, particularly gamma rays. Especially for light dark matter candidates, the predicted emission and the current constraints at around a GeV are comparable with those at around a keV.

In the right panel we consider models where dark matter pair-annihilates into $e^+e^-$ pairs. Such models, with very light masses of the order of an MeV, received a spur of interest as possible explanations to the excess detected by Integral-SPI from the galactic center region (see e.g. \cite{mevdm}). In the right panel, this time, we again fix $\langle\sigma v\rangle=3\times 10^{-26}$ cm$^3$/s, and tune the mass to 3 MeV (black solid), 10 MeV (black dashed), 1 GeV (blue solid) and 3 GeV (blue dashed). For the MeV dark matter models, the IC peak lies out of the scale we show, and for a conventional pair annihilation cross section, the final state radiation yield vastly exceeds both the hard X-ray and soft gamma-ray data on the flux of extragalactic background light. Notice that although the s-wave annihilation rate needed to explain the Integral-SPI data is indeed around the value we employ here, in several MeV dark matter models the dominant channel for thermal freeze-out in the early universe is through p-wave annihilation, and therefore doesn't contribute to dark matter annihilation today, at $T\simeq0$  \cite{mevdm}.

The GeV-scale mass models do feature the two peaks corresponding, respectively, to final state radiation and to the IC light. These models illustrate clearly that there is a strong complementarity between constraint on dark matter models from the gamma-ray and from the soft and hard X-ray extragalactic background. For instance, the strongest constraint on a 1-3 GeV mass dark matter particle annihilating into $e^\pm$ clearly comes from the extragalactic soft X-ray background, although a signal might also be expected in gamma rays.  These models are again ruled out for a conventional pair annihilation cross-section.

\section{Constraints on Dark Matter Models that account for the Pamela Positron Excess}\label{sec:pamela}

\begin{figure}[t]
\begin{center}
\mbox{\hspace*{-0.5cm}\includegraphics[width=8.cm,clip]{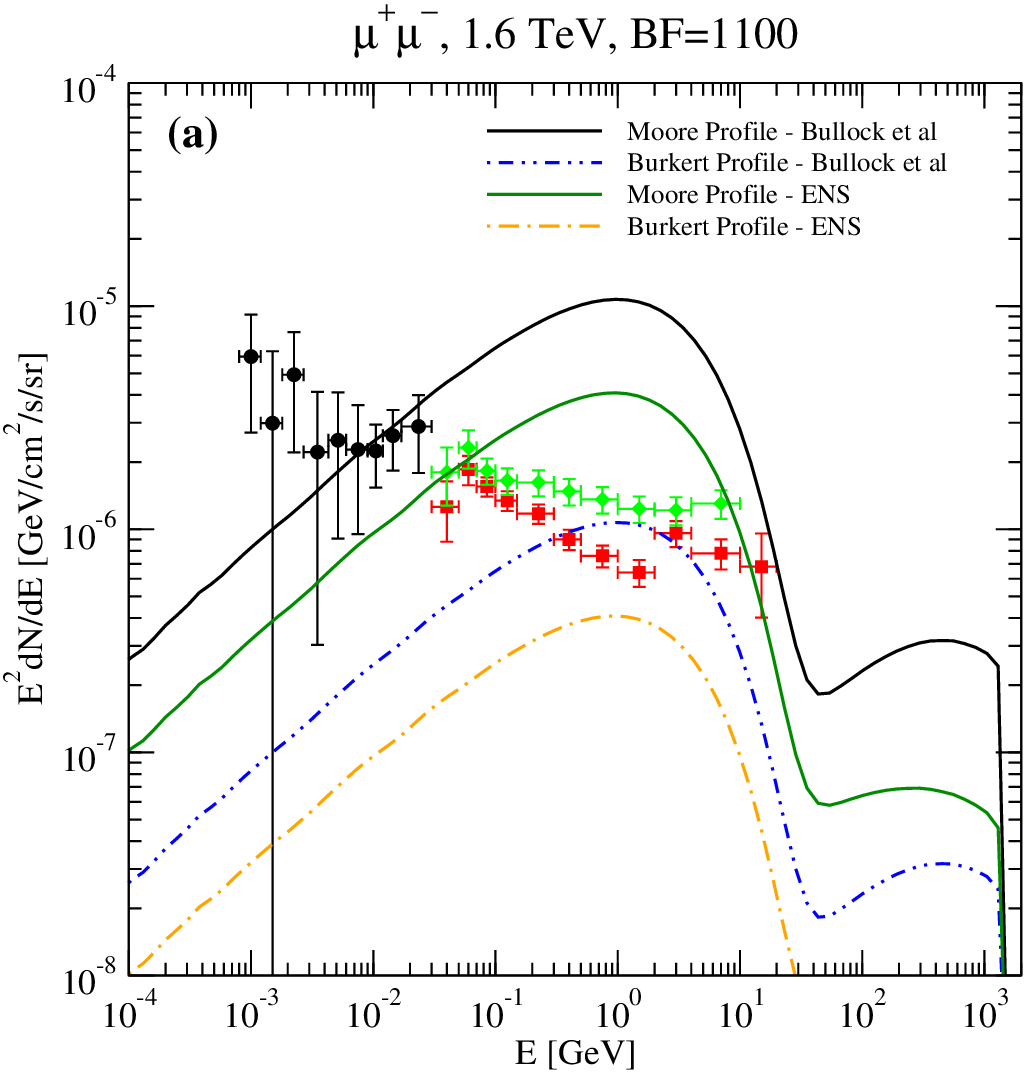} \qquad \includegraphics[width=8.cm,clip]{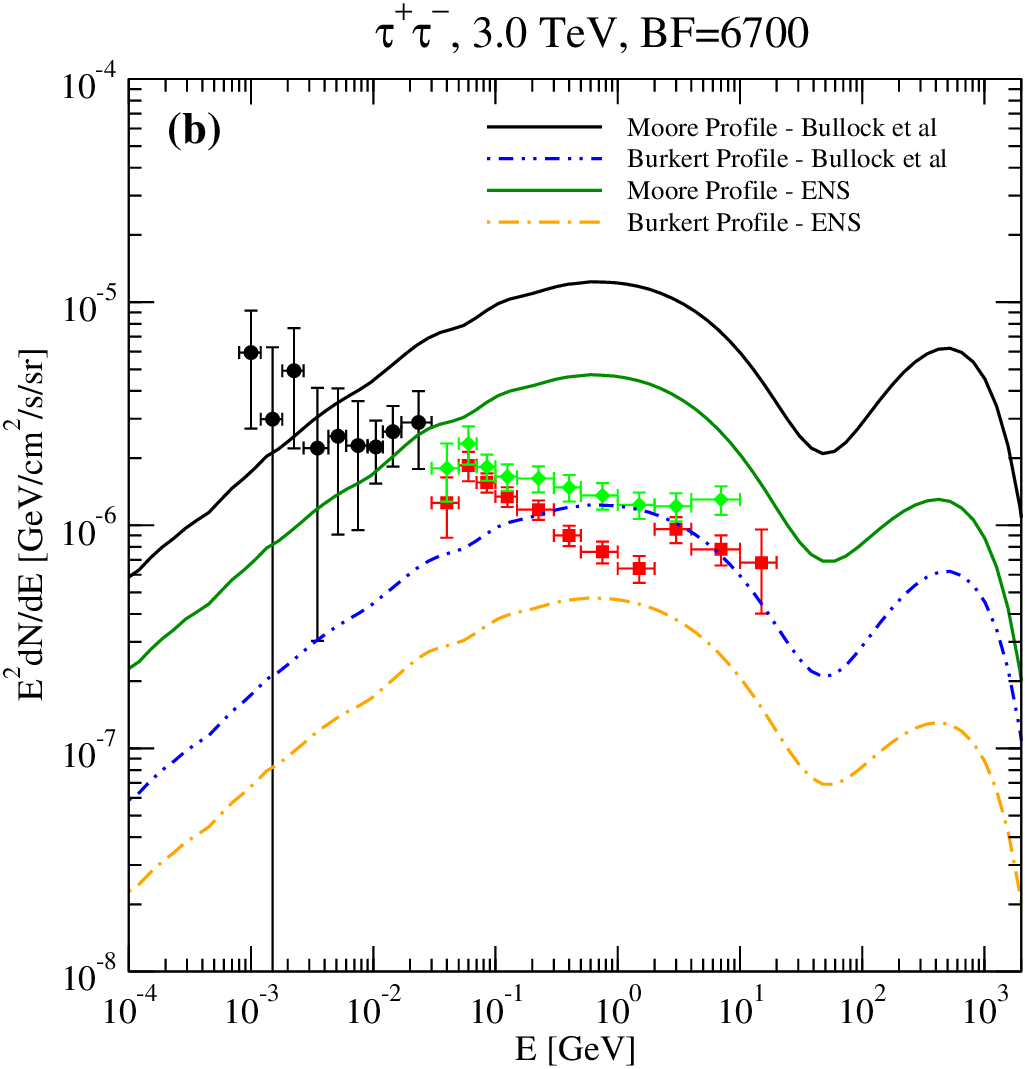}}\\[0.5cm]
\mbox{\hspace*{-0.5cm}\includegraphics[width=8.cm,clip]{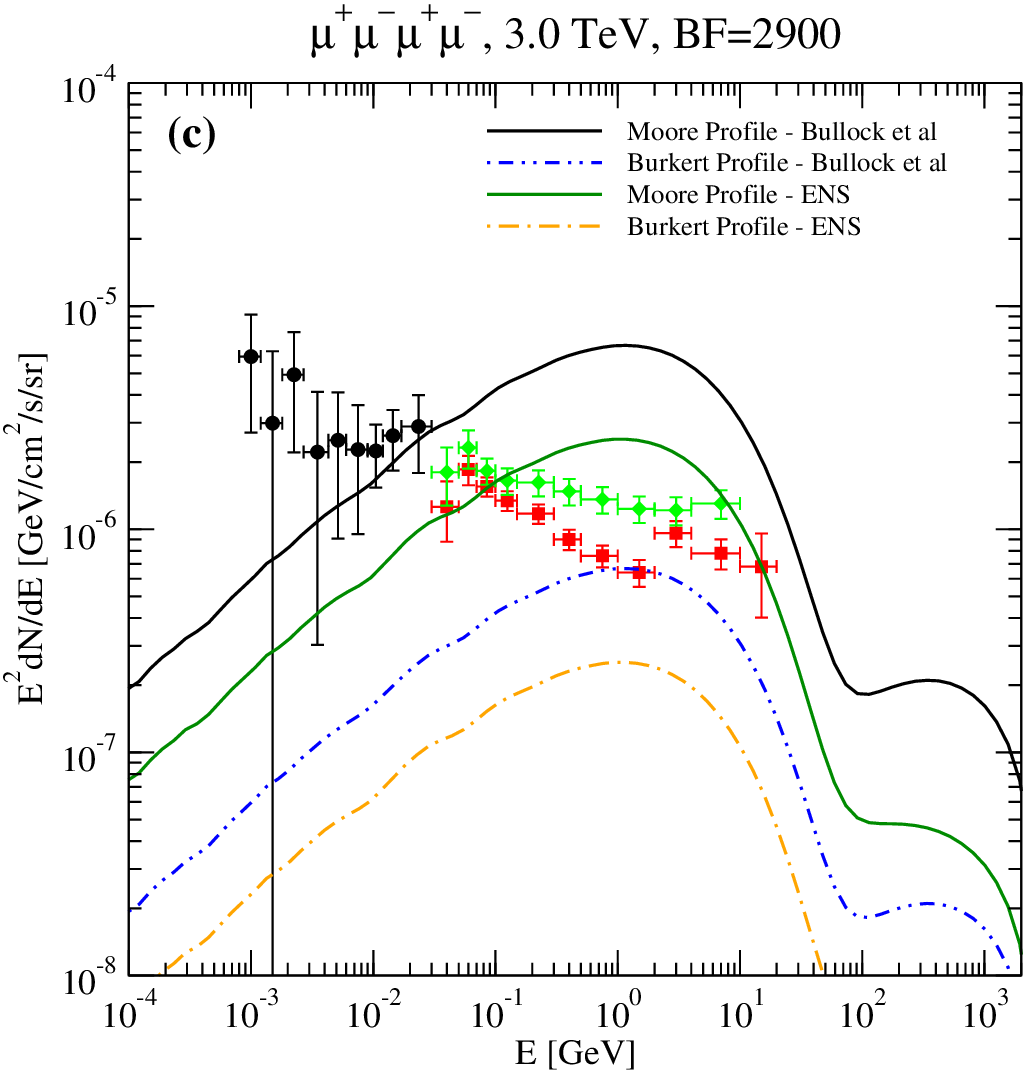} \qquad \includegraphics[width=8.cm,clip]{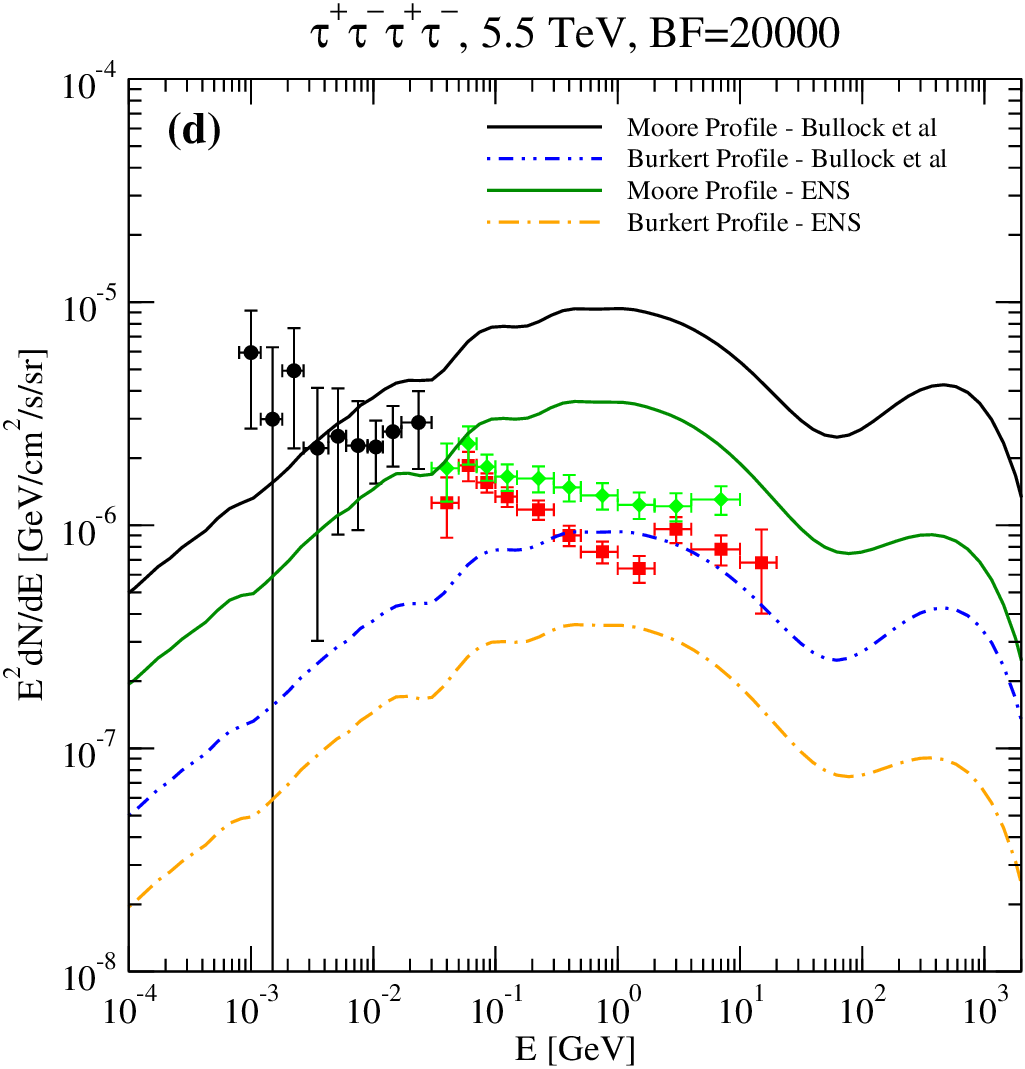}}

\caption{The dependence of  the dark matter all-redshift all-halo annihilation emission on the structure formation and halo model setup, for particle dark matter models that offer an explanation to the Pamela positron excess and that are compatible with the Fermi-LAT electron-positron data. Tab.~\ref{tab:models} gives the details of the mass, annihilation rate and final state for the four models we show in panels (a)-(d). \label{fig:pamela}
}
\end{center}
\end{figure}

In this section we specifically consider the all-redshift all-halo IC emission from models that have been recently invoked to explain the Pamela positron fraction data \cite{Adriani:2008zr}, and that are compatible with other experimental information, including Fermi-LAT data on the $e^\pm$ spectrum \cite{Abdo:2009zk}. We describe the benchmark models we consider here in tab.~\ref{tab:models}, and show the resulting all-redshift and all-halo emission from dark matter annihilation in the corresponding four panels of fig.~\ref{fig:pamela}, emphasizing the impact of different assumptions on the structure formation and halo density profile.

\begin{table}
\begin{center}
\begin{tabular}{|c|c|c|c|c|}
\hline
{\bf Panel} & mass/TeV & $\langle\sigma v\rangle/(3\times 10^{-26}$\  cm$^3$/s) & Final State & Ref.\\ 
\hline
{\bf (a)} & 1.6 & 1,100 & $\mu^+\mu^-$ & \cite{Bergstrom:2009fa}\\
{\bf (b)} & 3.0 & 6,700 & $\tau^+\tau^-$ & \cite{Meade:2009iu}\\
{\bf (c)} & 3.0 & 2,900 & $4\mu$ & \cite{Meade:2009iu}\\
{\bf (d)} & 5.5 & 20,000 & $4\tau$ & \cite{Meade:2009iu}\\
\hline
\end{tabular}
\end{center}
\caption{Details of the four benchmark models that could explain the Pamela and Fermi data, for which we calculate and show in fig.~\ref{fig:pamela} the all-redshift, all-halo annihilation emission.\label{tab:models}}
\end{table}

One such model, studied in Ref.~\cite{Bergstrom:2009fa}, has a mass of 1.6 TeV, an annihilation rate of $\langle\sigma v\rangle=3.3\times 10^{-23}$ cm$^3$/s and the dark matter annihilation proceeds into $\mu^+\mu^-$ pairs. Fig.~\ref{fig:pamela}, panel (a), shows the effect of changing our assumptions on $\Delta^2(z)$ from the benchmark setup employed in sec.~\ref{sec:results} (a Moore profile, and a Bullock et al structure formation setup, indicated with a black line). The double-dot dashed blue line shows the effect of assuming for all halos a cored inner profile (Burkert), while keeping the same structure formation setup as for the benchmark. EGRET data on the extragalactic gamma-ray flux still rule out this dark matter setup, even with this very conservative halo profile choice. An even stronger conclusion is reached by employing as an alternative the more conservative ENS structure formation setup: the solid green line shows how our benchmark setup is affected by this choice, which amounts to suppressing the IC emission by a factor 2-3, and the FSGR bump by an even larger factor (the ENS setup suppresses the contribution at low redshift compared to the Bullock et al setup). The only case where the IC emission from this dark matter model is (marginally) compatible with the EGRET data is by employing {\em both} a Burkert profile for all halos {\em and} the ENS structure formation setup (dot-dashed orange line). 

Panel (b) of fig.~\ref{fig:pamela} reinforces our conclusions for a slightly different dark matter setup, that Ref.~\cite{Meade:2009iu} pointed out to be one of the best scenarios to explain the Pamela positron excess in terms of dark matter annihilation. In this scenario, the dark matter particle mass is set to 3 TeV, the pair annihilation rate is $\langle\sigma v\rangle=2\times 10^{-22}$ cm$^3$/s and it proceeds into a pair of tau leptons. While the relative prominence of the IC peak is here less dramatic than in the $\mu^+\mu^-$ case, the cosmological IC emission firmly exceeds EGRET data on the GeV gamma-ray extragalactic background, unless the most conservative halo models and structure formation setups are invoked, and all other sources of extragalactic gamma rays are neglected. 

An additional possibility is that the dark matter pair-annihilates into a light particle $\phi$, which then decays into lepton pairs. This scenario has been shown to accomplish two tasks required for a viable dark matter interpretation of the Pamela data: (i) if $\phi$ is lighter than the proton mass, no antiprotons are produced in the dark matter annihilation even, in accord with the results of Pamela reported in \cite{pamelaantiproton} and (ii) the $\phi$ particle can be also responsible for a ``new force'' in the dark sector leading to a velocity-dependent enhancement that can explain the thermal relic abundance of dark matter as well as the large pair-annihilation rate in the Galaxy today \cite{ArkaniHamed:2008qn}.  We consider two examples of this class of models yielding cascading multiple lepton pairs (see e.g. \cite{nomurathaler}) in panels (c) and (d). The values of the masses and pair-annihilation rates for the $4\mu$ (c) and $4\tau$ channels are taken from Ref.~\cite{Meade:2009iu}. Although injecting softer $e^\pm$ pairs and therefore featuring a relatively flatter IC peak emission appearing at lower energies, these models do not generically escape the overproduction of extragalactic IC gamma ray photons.

\begin{figure}[t]
\begin{center}
\mbox{\hspace*{-0.5cm}\includegraphics[width=8.cm,clip]{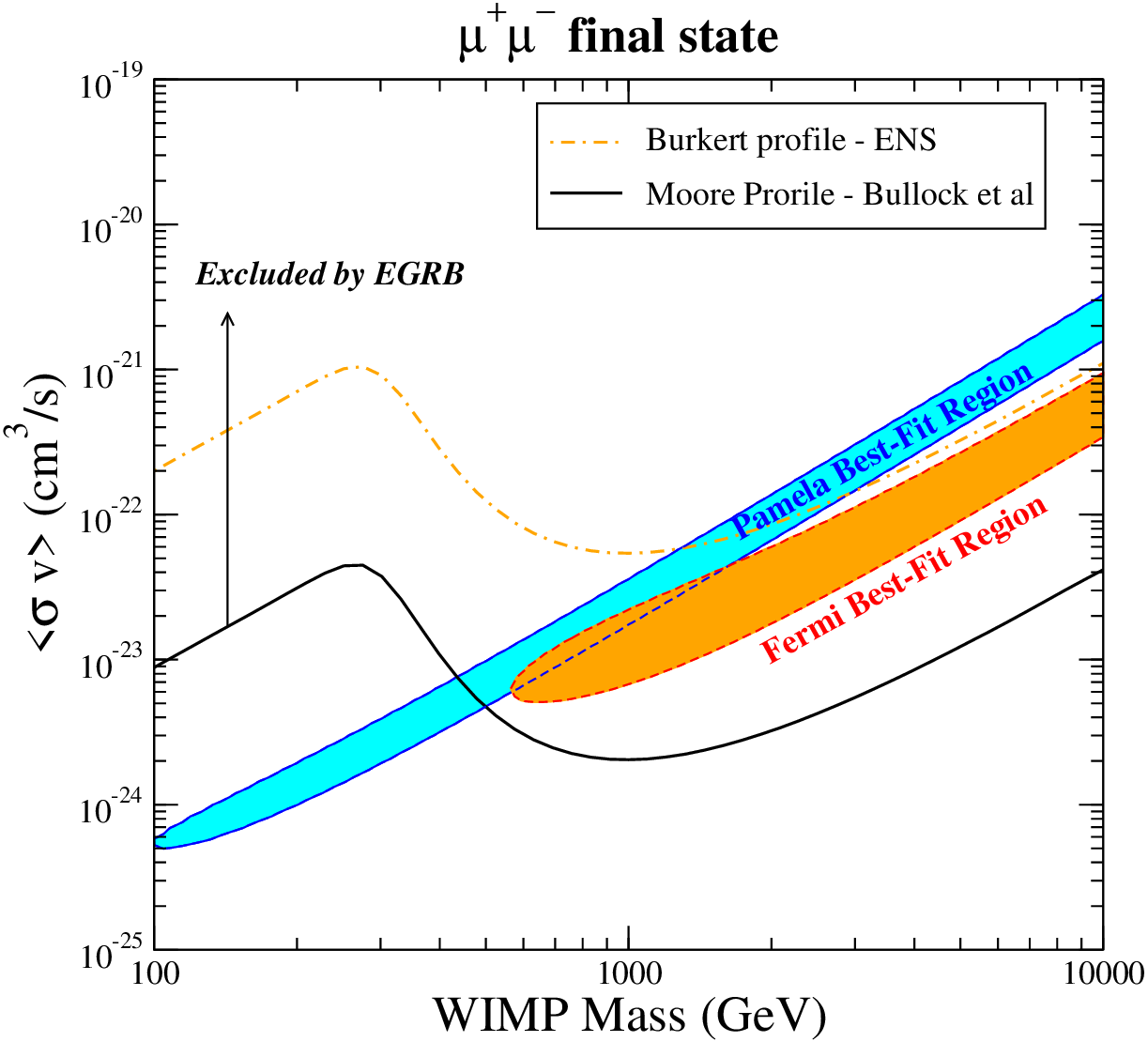} \qquad \includegraphics[width=8.cm,clip]{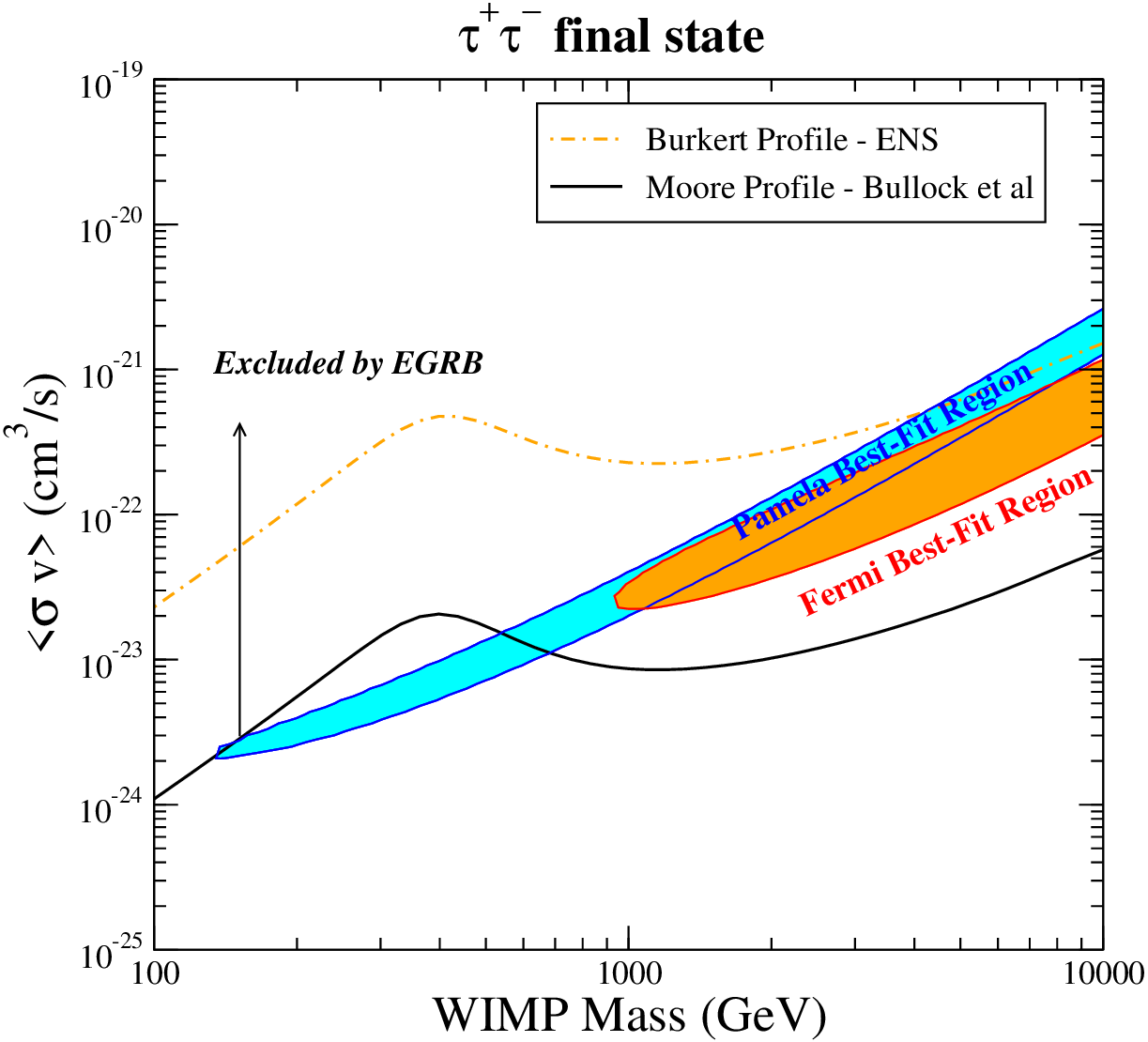}}
\caption{Constraints on the mass versus pair annihilation cross section parameter space plane from the all-redshift all-halo annihilation emission, for dark matter models pair-annihilating into $\mu^+\mu^-$ (left panel) and $\tau^+\tau^-$ (right panel). Points above the solid black line are excluded for a Bullock et al structure formation setup and a Moore halo profile, while those above the dot-dashed line are ruled out even with the more conservative ENS setup and Burkert halo profile. The shaded regions correspond to the best fit to the Pamela (cyan) and Fermi (orange) data, following the approach outlined in Ref.~\cite{Grasso:2009ma}.\label{fig:msv}}
\end{center}
\end{figure}
A summary of the constraints from the all-redshift all-halo annihilation emission on dark matter models that could explain the Pamela positron excess is given in fig.~\ref{fig:msv}. There we shade the regions of parameter space favored by a dark matter annihilation interpretation of the Pamela (cyan) and Fermi (orange) data, according to the procedure outlined in Ref.~\cite{Grasso:2009ma}, and we show the lines corresponding to the constraints from the extragalactic dark matter annihilation emission: Points above the solid black line are excluded for a Bullock et al structure formation setup and a Moore halo profile, while those above the dot-dashed line are ruled out even with the more conservative ENS setup and Burkert halo profile. Specifically, the constraints correspond to models that overproduce the extragalactic gamma-ray flux by more than 2-$\sigma$ for at least one of the EGRET bins. Notice that the shape of the curves representing the constraints reflects the fact that for low masses, final state radiation and gamma-rays from the prompt dark matter annihilation event over-ride the IC emission, which becomes dominant at masses larger than 300-400 GeV. For both final states the best fit region to both the Pamela and Fermi data is ruled out in the Bullock et al plus Moore setup, and even the most conservative scenario we consider here (ENS plus Burkert) rules out models with massive particles pair-annihilating into light leptons that could explain the Pamela positron anomaly.

Notice that the statement we make here that models that explain the Pamela excess are in tension with the all-redshift all-halo IC signal from dark matter annihilation is a particularly conservative one, given that in what we show above:

\begin{itemize}
\item we compared extragalactic background data with the emission from dark matter {\em only}, neglecting what are thought to be the most significant contributors to that extragalactic light, namely blazars (see e.g. \cite{Inoue:2008pk}); 
\item the only case where the predicted emission from the models we consider are not directly in tension with data is for (i) the most conservative dark matter halo profiles and (ii) a very conservative structure formation setup;
\item we neglected the IC contribution from background photons different from those in the CMB, which obviously enhances the total IC emission. Noticeable examples are starlight and dust re-scattered starlight as well as the UV and IR backgrounds;
\item electron-positron pairs produced in the gamma-ray absorption by the EBL can also, in principle, yield further emission (for instance from non-thermal bremsstrahlung in regions with high gas densities) that we have neglected, but that could add to the low-energy tail of the spectra we calculate here;
\item we neglected the galactic dark matter emission, both prompt and from IC. As recently shown in Ref.~\cite{Meade:2009iu} (see e.g. their fig.~4), the expected IC emission from annihilation in the galaxy for models that fit the Pamela and Fermi data is at the level of $E^2 {\rm d}N/{\rm dE}\sim 10^{-6}\ {\rm GeV}/{\rm cm}^2/s/sr$ for $E\sim1..100$ GeV, thus intermediate between the largest and smallest of the extragalactic signals we calculate here. The expected IC spectrum in the galactic emission case is harder than what we show here, due to the additional contribution of IC scattering off of starlight and dust-rescattered starlight. Although this galactic emission will not be entirely isotropic, and issues might arise in disentangling a diffuse galactic background model from a strictly extragalactic isotropic component, the additional contribution from dark matter annihilation in the galaxy definitely reinforces the constraints we outline in the present study;
\item finally, we neglected halo-dependent enhancement effects, including e.g. the so called Sommerfeld enhancement that could significantly boost the contribution from small structures with low velocity dispersions 
\end{itemize}

In short, we argue here that the IC emission from all redshifts provides a powerful probe to, and in many cases even rules out, the dark matter interpretation of the Pamela positron excess.

\section{Conclusions}

In this study we calculated the all-redshift, all-halo Inverse Compton emission from dark matter annihilation in addition to the corresponding final state gamma-ray emission. We outlined the available observational information and we articulated the dependence of that emission on both the dark matter particle model and on assumptions on the structure formation and halo density profiles. We argued that the unresolved extragalactic background light at X-ray frequencies provides a complementary handle on indirect dark matter detection to similar data at gamma-ray energies. In particular, the unresolved X-ray background rules out some low mass dark matter particle models with a conventional pair annihilation cross-section.  The Inverse Compton emission is particularly relevant for models where the dark matter pair-annihilates into leptonic final states, including those models which have been recently invoked to explain the positron anomaly reported by Pamela. We argued that those models are tightly constrained by the diffuse extragalactic light produced by the up-scattering of cosmic microwave background photons at all redshifts. In all but the most conservative halo profile and structure formation set-ups, these models overproduce the total extragalactic gamma-ray background, even before accounting for the contribution of blazars and other gamma-ray sources. These constraints will become increasingly powerful with the anticipated improvement in the understanding of the extragalactic background light expected soon with the Fermi-LAT data. The extragalactic diffuse emission is therefore yet another handle that Fermi-LAT provides to reveal the controversial origin of the anomalous excess of local high-energy positrons.

\section*{Acknowledgements}
T.E.J. is grateful for support from the Alexander F. Morrison Fellowship, administered through the University of California Observatories and the Regents of the University of California. S.P. is partly supported by an Outstanding Junior Investigator Award from the US Department of Energy, Office of Science, High Energy Physics and by Contract DEFG02-04ER41268, and 
by NASA Grant Number NNX08AV72G and NSF Grant PHY-0757911.

\end{document}